\documentclass[fleqn,usenatbib]{mnras}

\usepackage{newtxtext,newtxmath}

\usepackage[T1]{fontenc}
\usepackage{ae,aecompl}

\usepackage{graphicx}	
\usepackage{amsmath}	
\usepackage{amssymb}	
\usepackage{xcolor}

\newcommand{\teff}{$\sc {T_{eff}}$}
\newcommand{\logg}{log(g)}

\newcommand{\beix}{$^{9}$Be}
\newcommand{\bevii}{$^{7}$Be}
\newcommand{\beixii}{$^{9}$Be\,{\sc ii}}

\newcommand{\livii}{$^{7}$Li}
\newcommand{\livi}{$^{6}$Li}

\newcommand{\iiihe}{$^{3}$He}
\newcommand{\ivhe}{$^{4}$He}

\title[Lithium and Beryllium in the GE Galaxy]{Lithium and Beryllium in the Gaia-Enceladus Galaxy}

\author[P. Molaro et al.]
{P. Molaro,$^{1,3}$\thanks{E-mail: paolo.molaro@inaf.it}
G. Cescutti,$^{1}$
and X. Fu$^2$
\\
$^1$ INAF, Osservatorio Astronomico di Trieste, Via G.B. Tiepolo 11, I-34143 Trieste, Italy\\
$^2$ The Kavli Institute for Astronomy and Astrophysics at Peking University, Beijing 100871, China\\
$^3$ IFPU, Istitute for the Fundamental Physics of the Universe, Via Beirut,  2, 34151, Grignano, Trieste, Italy}

\date{Accepted XXX. Received YYY; in original form ZZZ}

\pubyear{2015}

\begin{document}
\label{firstpage}
\pagerange{\pageref{firstpage}--\pageref{lastpage}}

\maketitle

\begin{abstract}
Data from   Gaia DR2 and APOGEE surveys  
revealed a relatively new component  in the inner Galactic halo,  which is likely   the dynamical remnant  of  a disrupted 
dwarf galaxy named Gaia-Enceladus  that collided  with the Milky Way about 10 Gyrs ago. This merging event offers an extraordinary opportunity to study  chemical abundances of  elements in a dwarf galaxy, since they   are generally  hampered in  external galaxies. Here, we focus on \livii\ and \beix\ in dwarf stars which are  out of reach even in  Local Group galaxies.
 Searching in   GALAH, Gaia-ESO survey and in  literature, we found several  existing \livii\ abundance determinations of stars belonging to the Gaia-Enceladus galaxy. The \livii\ abundances of stars at the low metallicity end overlap  with those of the   Galactic halo.
 These are  effective   extragalactic \livii\ measurements,  which suggest  that the \livii\ {\it Spite plateau}   is universal, as is the   cosmological \livii\ problem.
 We found a  \livii-rich giant out of 101 stars, which   suggests a small percentage   similar  to that of the Milky Way.  We also collect   \beix\ abundance for a subsample of  25  Gaia-Enceladus stars from  literature. Their abundances share the Galactic [Be/H]  values at the low metallicity end but grow slower with  [Fe/H] and show  a reduced  dispersion. This suggests that   the scatter   observed in the Milky Way could reflect the different \beix\ evolution patterns of   different stellar components which are mixed-up in the Galactic halo.

\end{abstract}




\section{Introduction}

In the last decades kinematical and chemical surveys of the stars of the Galactic Halo   revealed   streams and structures  belonging to different stellar groups \citep{apogee2012,gaia2016}.
 One of the  most studied   is  Sagittarius dSph,  \citep{Ibata2001}. The stellar component  of the Galactic halo  showing   low-[$\alpha$/Fe]  probably was also  accreted from local dwarf galaxies   \citep{gratton2003,brook2003,Nissen2010,Kirby2009,Tolstoy2009}.
More recently, the ground based  spectrographic survey APOGEE together with the astrometric results from the Gaia satellite DR2   revealed a  component in the inner halo which shows  a distinctive   motion 
and   metallicities $Z \approx Z_{\sun}/10$ which are relatively more metal-rich than   the Galactic halo  \citep{Helmi2018,Haywood2018,Belokurov2018}. 
This new structure,  called Gaia-Enceladus or  Gaia Sausage,   likely represents  a  disrupted dwarf galaxy  after collision  with the  Milky Way   about  $10\,\text{Gyr}$ ago.  

 The stellar remnants of this merging  offer  unique  opportunities to study in detail the  abundances of the elements in a dwarf galaxy, normally  hampered by their large distance. 
 A detailed chemical analysis of the most studied elements in these stars has been already performed in \citep{Vincenzo2019}. Here, we focus on the elements \livii\ and \beix\  which have special nucleosynthetic origin and  about which nothing or very  little is known in  extragalactic environments.

The nucleosynthetic origin of the light elements Li, Be and B  differs from that of the other  chemical 
elements. As was first recognized by \cite{Burbidge1957} they cannot be made in the stellar interior or in the explosive phases.  They suggested an {\it X-mechanism}  likely connected to spallation processes taking place  onto the surfaces of magnetic stars or somewhat  in  the  supernovae blow out. In fact, Li undergoes   multiple  nucleosynthetic processes. \citet{Wagoner1967} have shown that \livii\ is made in the primordial nucleosynthesis and \citet{Reeves1970} suggested a  cosmic-ray spallation occurring
in the interstellar medium.  The 
 Li/H abundance measurements in the old Galactic halo stars are
a factor of 3.5 lower than  primordial nucleosynthesis predictions \citep{Fields2014} and it is not clear 
whether this mismatch comes
from  uncertainties in stellar astrophysics or nuclear inputs, or whether
there is  new physics at work, see    \citet{Fields2011} and references therein. Cross
sections of  BBN reactions are constrained by extensive laboratory
measurements therefore making   nuclear fix increasingly unlikely.
Another  possibility   is a  \livii\  destruction over the long stellar lifetimes  by  mechanisms such as  microscopic diffusion \citep{Korn2006}, rotational
mixing \citep{Pinsonneault1998}, or pre-main-sequence depletion \citep{Fu2015}.
However,    fine tuning of the initial stellar parameters is required to reduce lithium to the observed  levels. On the other hand,
one or more  Galactic Li sources are  needed to increase the  Li abundance from the primordial value to that  presently observed either in  meteorites  or in  young stars of the Galaxy. Quite recently, novae  have  gained favour  as the probable main Galactic sources with  detection of \bevii\, which later decays into \livii\, in their outburts \citep{Tajitsu2015, Molaro2016, Izzo2018, Izzo2018b, Cescutti2019}.
 $^{9}$Be is the  only   long-lived isotope of beryllium and  is a pure product of cosmic-ray spallation \citep{Reeves1970}.  Early theoretical models of the Galaxy predicted  a secondary behaviour with a  quadratic dependence     on metallicity. However, early measurements   of Be and B in metal-poor stars revealed  a  linear relation with metallicity which is characteristic of a primary production \citep{Rebolo1988,Duncan1992}.  
\citet{Duncan1992}   suggested that the principal channel of synthesis involves the collision of cosmic-ray CNO nuclei from the supernovae  with interstellar protons as was already envisaged by \citet{Burbidge1957}.   Being a primary element,    \beix\ abundance should trace the chemical evolution of other primary elements  belonging  Galactic component. 
 In this paper we carefully search for  extant \livii\ and \beix\ abundances of stars belonging to   Gaia-Enceladus and we seek  signatures of a different chemical evolution by comparing them to those of the Galaxy.

\section{Lithium abundances  in Gaia-Enceladus}

\citet{Helmi2018} provided a sample of 4644  suggested  Gaia-Enceladus  member stars,  a subsample  of which are  in the catalogue of  The Apache Point Observatory Galactic Evolution Experiment (APOGEE)  with   determined abundances for 18 elements \citep{apogee2012}.
The Gaia-Enceladus subsample with APOGEE measurements includes  stars with   $[\alpha/\text{Fe}]$ values lower than thatof  the  Milky Way halo stars in the metallicity range $-$1.5<[Fe/H]<$-$0.5, which are 
  typical of dwarf spheroidal galaxies and are also observed in the  Damped Lyman- $\alpha$ galaxies  \citet{Rafelski2012}.
 APOGEE does not provide \livii\ abundances. Thus, we searched  for the Gaia-Enceladus component in the GALAH DR2 survey \citep{Buder18}, the Gaia-ESO DR3 survey \citep{gaiaeso2012} and in the literature by cross matching the larger sample of Gaia-Enceladus stars provided by  \citet{Helmi2018} by considering their selection criteria, namely a  distance of $<$ 5 kpc and Lz $<$ 150.

The cross-match between  Gaia-Enceladus candidates  with the  GALAH   yielded  121 stars  which are provided in the online Table.  The selection has been restricted to GALAH stars with  flag=0 for the stellar parameters. 11 stars out of this sample
have  [Fe/H]$<$ -0.8, Teff $>$ 5700 K and Logg $>$ 3.65. The cuts have been chosen to avoid the stars in which  Li has been depleted or diluted by  main sequence or post main-sequence evolution.
 The Gaia-ESO survey provided   only one giant star.   By cross matching with the literature databases of  JINA  \citep{JINA},  SAGA  \citep{SAGA} and  \citet{Aguilera18} we found 31 additional Gaia-Enceladus   dwarf star candidates that match the same criteria  with extant lithium abundances \footnote{A(Li) = $\log \frac{n(Li)}{n(H)}+12$}.  All 43 Gaia-Enceladus candidates are listed in Table \ref{tab:literature}. The kinematical properties of the selected Gaia-Enceladus candidates are reported in Table \ref{tab:cinematica}. 
  
 The energy (En) and the  angular momentum in the Z direction  ($L_Z$)  are adopted from  \citet{Helmi2018}. 
In addition, we have also computed the orbital parameters,
 including the apocenter distance ($r_{apo}$), the pericenter distance ($r_{peri}$),  the maximum distance from the Galactic plane ($Z_{max}$) and   eccentricity (ecc), 
 based on the stellar orbit in the last 1 Gyr. 
 For this  calculation we use the public licensed code \texttt{GALPOT} following the method described in \citet{galpot}, and assume a Galactic potential which includes thin and thick stellar disks, bulge, halo, and a gas disk. 
 To note that almost all  selected Gaia-Enceladus stars show high, almost parabolic eccentricity,   which is different from the nomal halo stars with a more circular orbit.  
 This further  supports the hypothesis that  they are  accreted. 
 In the selection for  Gaia-Enceladus candidates  there is a strong probability of contaminations by thick disk stars in particular at  relatively high metallicity.
 For instance,  G 5-40  satisfies the kinematical cuts but shows a too high  [$\alpha$/Fe] ratio for a  Gaia-Enceladus member  \citep{Nissen2010}. 
 In the following,  we provide  an analysis  both with and without this star. 
 
 \begin{figure}
\includegraphics[width=\columnwidth]{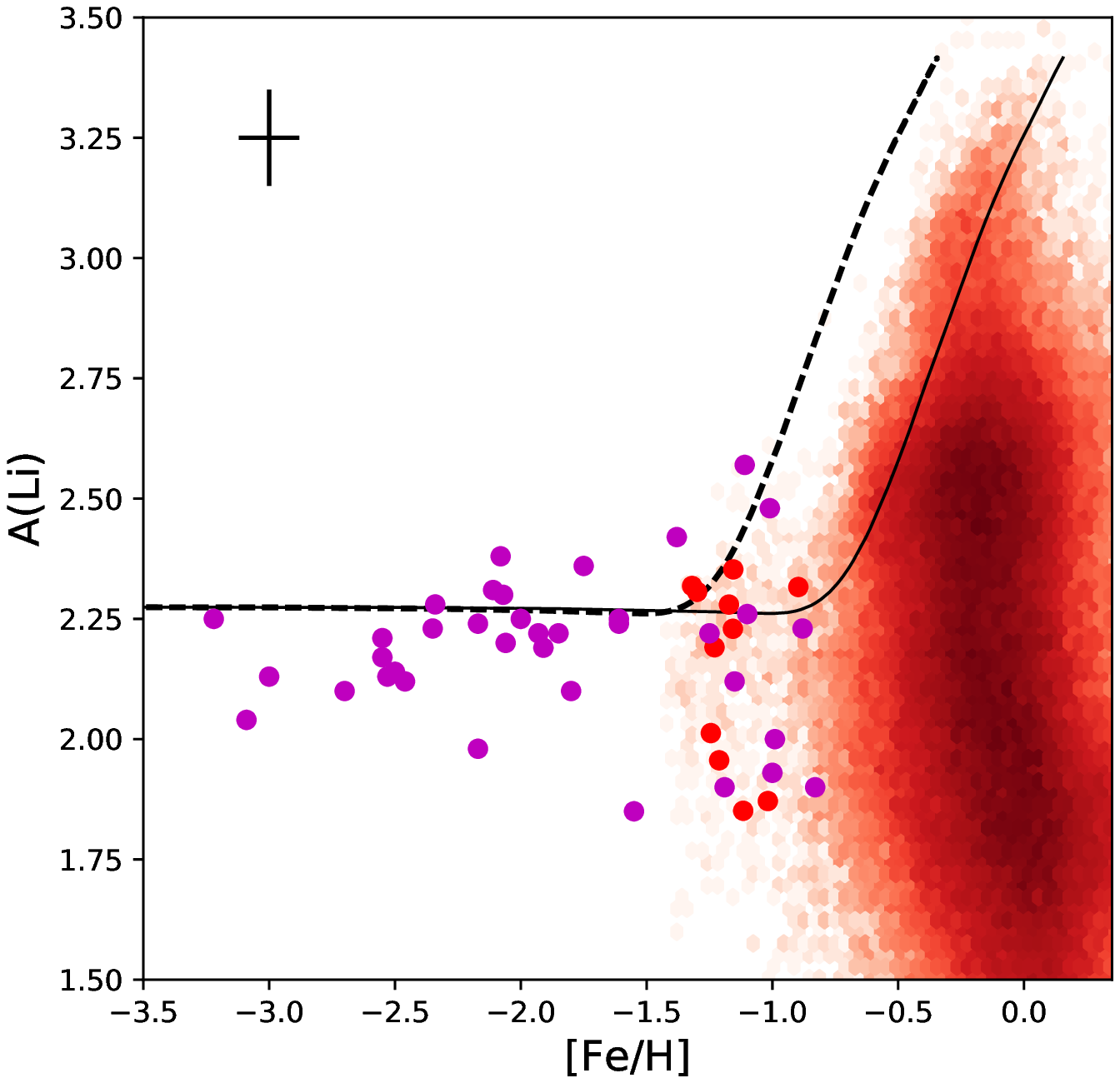}
\caption{  Li abundances  A(Li) as a function of  [Fe/H]  of all dwarf Gaia-Enceladus candidates with 
(logg$>$3.65, $T_{eff}$ $>$ 5700 K) found in the GALAH survey, red dots, and in literature, magenta, listed in Table \ref{tab:literature}. The GALAH   stars have  recommended stellar parameters
(flag cannon=0) \livii\ abundance.  The cross in the top left corner shows  a typical $\pm \sigma$ abundance error. The best model of the  Li evolution for the thin disc  presented in \citet{Cescutti2019} is shown in  black solid line. With a black dashed line is the same model but   shifted by -0.5 dex in metallicity as a proxy for  the time delay  expected  for a dwarf galaxy as Gaia-Enceladus.}
\label{Li_ence}
\end{figure}

\begin{table*}
 \label{tab:literature}
 \caption{Gaia-Enceladus  candidate stars with Li and Be abundances measured  from literature with  $\logg  >$ 3.65 and $T_{eff}>$ 5700 K.  References for the stellar parameters and  Li abundance  are reported  in the last column (first number) and those  for Be in the second number: 1 \citep{Buder18}, 2 \citet{Fulbright2000}, 3 \citet{Boesgaard2006}, 4 \citet{Boesgaard2005}, 5 \citet{Smiljanic2009}, 6 \citet{Charbonnel2005}, 7 \citet{Asplund2006}, 8 \citet{Spite2015}, 9 \citet{Siqueira2015}, 10 \citet{Ramirez2012}, 11 \citet{Delgadomena2015}, 12 \citet{Delgadomena2014}, 13 \citet{placco2016}, 14 \citet{bonifacio1997}, 15 \citet{Boesgaard2011} 16 \citet{Smiljanic2009}, 17 \citet{Rich2009}, 18 \citet{tan2009}, 19 \citet{Roederer2014}, 20 \citet{melendez2010}, 21 \citet{Charbonnel2000}, 22 \citet{hosford2009} \citet{Aguilera18}.  The oxygen abundances are from \citet{Boesgaard2011} or \citet{Smiljanic2009}, 
 in this latter  case [O/H] is inferred by their [$\alpha$ -element/Fe].
 }
\begin{tabular}{|r|l|r|r|r|r|r|r|r|r|r|l|}
\hline
  \multicolumn{1}{|c|}{Gaia source\_id} &
  \multicolumn{1}{c|}{Name} &
  \multicolumn{1}{c|}{\teff} &
  \multicolumn{1}{c|}{\logg} &
  \multicolumn{1}{c|}{[Fe/H]} &
  \multicolumn{1}{c|}{[O/H]} &
  \multicolumn{1}{c|}{A(Li)} &
  \multicolumn{1}{c|}{[Fe/H]$_{Be}$} &
  \multicolumn{1}{c|}{$\epsilon _{Fe}$} &
  \multicolumn{1}{c|}{A(Be)} &
  \multicolumn{1}{c|}{$\epsilon _{Be}$} &
  \multicolumn{1}{c|}{ref} \\
\hline
  6086864760409366656 & TYC 8248-1737-1 & 5901 & 4.05 & -0.89 &  & 2.31 &      &      &  &      & 1\\
  4725550450463451904 & L  126-11 & 5857 & 3.79 & -1.02 &  & 1.87 &      &      &  &      & 1\\
  5459976109889190144 & TYC 7174-224-1 & 5929 & 4.02 & -1.12 &  & 1.85 &      &      &  &      & 1\\
  5750434405835685888 &  & 5834 & 3.84 & -1.16 &  & 2.35 &      &      &  &      & 1\\
  6679323239394561792 &  & 5888 & 4.06 & -1.16 &  & 2.23 &      &      &  &      & 1\\
  5946574193490564480 &  & 5982 & 3.83 & -1.17 &  & 2.28 &      &      &  &      & 1\\
  5781595596159463040 & TYC 9429-2667-1 & 5935 & 3.92 & -1.21 &  & 1.95 &      &      &  &      & 1\\
  6383892436469819008 &  & 5740 & 4.05 & -1.23 &  & 2.19 &      &      &  &      & 1\\
  6729270234418615552 & CD-38 13129 & 5956 & 3.96 & -1.24 &  & 2.01 &      &      &  &      & 1\\
  5242632244811706496 & TYC 9213-2091-1 & 5986 & 3.9 & -1.3 &  & 2.3 &      &      &  &      & 1\\
  3155410389590889856 & G  89-14 & 5834 & 3.8 & -1.32 &  & 2.32 &      &      &  &      & 1\\
  32655224762711936 & G4-36 & 5810 & 3.7 & -2.17 &  & 1.98 &      &      &  &      & 19\\
  3846427888295815552 & HE0938+0114 & 6030 & 3.7 & -3.09 & -2.5 & 2.04 & -2.67 & 0.11 & -1.0 & 0.12 & 19,17\\
  866863321051682176 & BD+24 1676 & 6140 & 3.8 & -2.7 & -1.94 & 2.1 & -2.55 & 0.09 & -1.28 & 0.12 & 19,15\\
  1289512635833404032 & G166-47 & 5960 & 3.7 & -2.46 &  & 2.12 &      &      &  &      & 19,15\\
  4761346872572913408 & HIP 24316 & 5725 & 4.4 & -1.5 & -1.33 & 2.12 & -1.5 & 0.15 & 0.04 & 0.13 & \\
  5181063205724188032 & G75-56 & 6190 & 3.9 & -2.35 & -1.74 & 2.23 & -2.38 & 0.08 & -0.84 & 0.12 & 19,15\\
  1776289248313154688 & BD+17 4708 & 6025 & 4.0 & -1.61 & -1.09 & 2.25 & -1.5 & 0.15 & -0.34 & 0.13 & 2,15\\
   2658240166703766016 & BD+02 4651 & 6100 & 3.8 & -1.75 & -1.18 & 2.36 & -1.9 & 0.09 & -0.58 & 0.12 & 2,15\\
  4272653983123701120 & G21-22 & 5916 & 4.6 & -1.01 & -1.02 & 2.48 & -1.02 & 0.25 & 0.33 & 0.16 & 3,15\\
  2910503176753011840 & LTT2415.00 & 6295 & 4.1 & -2.11 &  & 2.31 &      &      &  &      & 4\\
  125750427611380480 & G37-37 & 5990 & 3.8 & -2.34 &  & 2.28 &      &      &  &      & 4\\
  29331710349509376 & G05-19 & 5942 & 4.2 & -1.1 & -0.62 & 2.26 & -1.1 & 0.15 & -0.01 & 0.13 & 5,15\\
  2905773322545989760 & HIP 25659 & 5855 & 4.5 & -2.0 &  & 2.25 &      &      &  &      & 6\\
  5586241315104190848 & HD59392 & 5936 & 4.0 & -1.61 &  & 2.24 &      &      &  &      & 7\\
  5551565291043498496 & CD-48 2445 & 6222 & 4.3 & -1.93 &  & 2.22 &      &      &  &      & 7\\
  949652698331943552 & G107-50 & 6030 & 3.9 & -2.06 &  & 2.2 &      &      &  &      & 19\\
  5617037433203876224 & W 0725-2351 & 6050 & 4.2 & -2.55 &  & 2.17 &      &      &  &      & 8\\
  870628736060892800 & G88-10 & 6033 & 4.2 & -2.53 & -1.86 & 2.13 & -2.61 & 0.07 & -1.08 & 0.12 & 6,15\\
  1097488908634778496 & G234-28 & 5870 & 3.8 & -1.8 &  & 2.1 &      &      &  &      & 19\\
  6268770373590148224 & HD 140283 & 5750 & 3.7 & -2.5 & -1.72 & 2.14 & -2.41 & 0.08 & -0.94 & 0.14 & 9,15\\
   731253779217024640 & HIP 52771 & 5937 & 4.5 & -1.85 & -1.4 & 2.22 & -1.8 & 0.15 & -0.55 & 0.13 & 10,16\\
   791249665893533568& BD+51  1696 & 5725 & 4.6 & -1.19& -0.53& 1.9 &-1.21& &-0.33&&10,15 \\
  2279933915356255232 & BD+75   839 & 5770 & 4.0 & -0.99 &  & 2.0 &      &      &  &      & 10\\
  2427069874188580480 & HIP 3026 & 6223 & 4.2 & -1.11 & -0.81 & 2.57 & -1.2 & 0.15 & -0.11 & 0.13 & 10\\
  5486881507314450816 & HIP 34285 & 5928 & 4.3 & -0.88 & -0.25 & 2.23 & -0.9 & 0.15 & 0.15 & 0.13 & 11,16\\
   6859076107589173120 & HIP 100568 & 5801 & 4.6 & -1.0 & -0.65 & 1.93 & -1.0 & 0.15 & 0.08 & 0.13 & 12,16\\
  
  4468185319917050240 &  G 20-24 & 6190 & 3.9 & -1.91 & -1.41 & 2.19 & -1.92 & 0.08 & -0.72 & 0.17 & 18,16\\
  4376174445988280576 & BD +2 3375 & 5800 & 4.1 & -2.39 & -1.48 &  & -2.39 & 0.17 & -0.74 & 0.15 & 17,17\\
   588856788129452160 & BD 9 2190 & 6008 & 3.9 & -3.0 & -2.38 & 2.13 & -3.0 & 0.09 & -1.22 & 0.11 & 7,17\\
  61382470003648896 & G 5 -40 & 5863 & 4.2 & -0.83 & -0.60 & 1.9 & -0.83 & 0.15 & 0.85 & 0.13 & 16,16 \\
  4715919175280799616 & HIP 7459 & 5909 & 4.46 & -1.15 & -0.98 & 2.12 & -1.15 & 0.15 & 0.12 & 0.13 & 16,16\\
  3643857920443831168 & G64-37 & 6300 & 4.2 & -3.22 & -2.32 & 2.25 & -3.28 & 0.05 & -1.4 & 0.11 & 13,15\\
  5709390701922940416 & HIP 42592 & 6040 & 4.1 & -2.17 & -1.56 & 2.24 & -2.0 & 0.15 & -0.58 & 0.13 & 14,16\\
  5184824046591678848 & LP 651-4 & 6030 & 4.3 & -2.89 & -2.04 &  & -2.89 & 0.08 & -1.12 & 0.12 & 17,17\\
  761871677268717952 & BD 36 2165 & 6315 & 4.3 & -1.38 &  & 2.42 &      &      &  &      & 20\\
   2722849325377392384 & BD 7 4841 & 5922 & 3.9 & -1.25 &  & 2.22 &      &      &  &      & 21\\
  16730924842529024 & BD 11 468 & 5739 & 4.6 & -1.55 &  & 1.85 &      &      &  &      & 20\\
  1458016709798909952 & BD 34 2476 & 6416 & 4.0 & -2.07 &  & 2.3 &      &      &  &      & 20\\
  5806792348219626624 & CD -71 1234 & 6194 & 4.5 & -2.55 &  & 2.21 &      &      &  &      & 22\\
  3699174968912810624 & HE1208-0040 & 6304 & 4.3 & -2.08 &  & 2.38 &      &      &  &      & 20\\
  4228176122142169600 & G 24-25 & 5752 & 3.7 & -1.56 & -0.98 &  & -1.56 & 0.09 & -0.73 & 0.12 & 15,15\\
  5133305707717726464 & BD -17 484 & 6110 & 3.6 & -1.56 & -0.95 &  & -1.56 & 0.09 & -0.37 & 0.12 & 15,15\\
\hline\end{tabular}
\end{table*}

 The A(Li) abundances vs. [Fe/H] of the selected stars are shown in Figure \ref{Li_ence}. From the figure it can be seen that  the Gaia-Enceladus stars show a very similar    behavior  to stars in the Milky Way in particular at low metallicity.   The mean value of the 17 stars with [Fe/H] $<$ 2.0 in Table \ref{tab:literature} is A(Li)=2.18$\pm 0.10$ to be compared with the A(Li)= 2.199$\pm0.086$ found by \citet{Sbordone2010} in a similar metallicity range.  However, there are   three   stars which present an enhancement of lithium at [Fe/H] $\approx$ -1 which suggests a slightly different Li evolution  in this dwarf galaxy. If novae are the main source which drive the Li enrichment in the Galaxy in a dwarf galaxy which is characterised by a slower star formation rate, their effects should start to be evident at lower metallicity. To  sketch the evolution of Li  in Gaia-Enceladus, we show in Figure \ref{Li_ence} the thin disk evolution (thin line) and the same results with an offset of $-$0.5 dex in metallicity; the offset is applied to mimic the typical lower efficiency in a satellite galaxy \citep{Matteucci90}.  A  possible model for the Li evolution in the Gaia-Enceladus galaxy has been  also  discussed  in  \citet{cescutti2020b}.

  \section{Li-rich  giants in Gaia-Enceladus}  
  
\begin{figure}
	\includegraphics[width=\columnwidth]{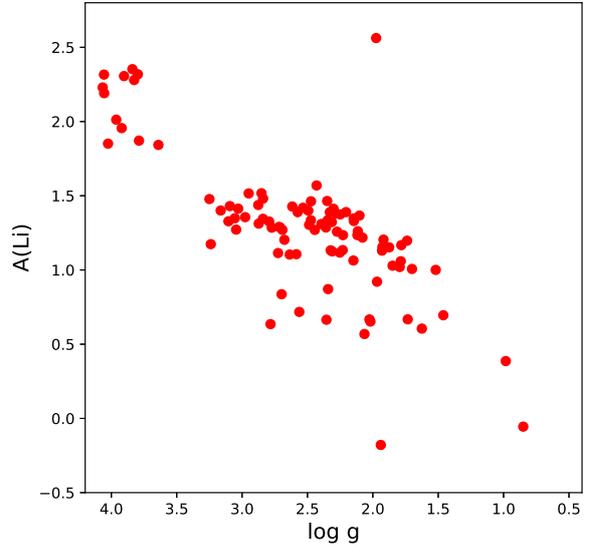}
    \caption{
    \livii\ abundances of all the Gaia-Enceladus star candidates from the GALAH survey.}
    \label{fig:giant}
    \end{figure}

When a star evolves off the main sequence, the surface convective zone deepens and material from the hotter interior is dredged up to the surface. 
Since \livii\ is a fragile element that is efficiently burned at a temperature of several millions  degrees,  both a dilution with the \livii\-free hot interior material and some \livii\ burning at the bottom of the surface convective zone  make the \livii\ abundance in a giant star  decrease by $\sim$1.3 - 1.5 dex below its main sequence value. 
Some extra mixing after the RGB bump reduces the surface \livii\ abundance to an even lower value \citep{Charbonnel2007}.  
The A(Li) vs. $\log g$ evolution of the Gaia-Enceladus stars is shown in Figure \ref{fig:giant} where the $\log g$ is considered an index of the evolutionary phase. The  giant stars show the characteristic depletion   flattening at A(Li) $\approx$  1.3 and with a minor fraction showing the sign of extra-mixing with almost no \livii\ after $log g \approx 1.8$.  
A(Li) measured in  red giants is in  good agreement with the results for the Galactic halo field, e.g.  \citet{Mucciarelli2012} found an average of A(Li)=0.97 for the Milky Way halo stars. 

In the Galaxy there is a small fraction of  giants with relatively large A(Li) abundances   $\gtrapprox$ 2.0, i.e. the Li-rich giants. This lithium could be produced  if there were extra mixing by a Cameron-Fowler mechanism, which requires some \bevii\ produced in the stellar interior and transported to the stellar surface by convection where it decays into \livii\ \citep{CameronFowler1971,sackmann1999}. Another possibility is that  Li has been preserved instead of undergoing post main-sequence dilution. The precise mechanism is controversial and it is not clear if there is net production of \livii\ or merely a preservation of the initial one \citep{Casey2016}.
Several recent giant stars have been detected   \citet[e.g.][]{Li2018,Yan2018, Smiljanic2018, gao2019}, including one with very high abundance.  However, this star is still in the main sequence or has just left it, and its high abundance is quite anomalous. 
 It is therefore  interesting to see whether   Li-rich giants are present   in other galaxies \citep[e.g.][]{kirby2012}.  In our sample of 121 Gaia-Enceladus candidates drawn from GALAH  and  shown in Figure \ref{fig:giant}, there is  one star 15344465-3331196 (T$_{eff}$ =4837 K, log=1.97, [Fe/H]= -0.37) with A(Li) $\approx$ 2.6, out of 101 stars  with log$\le$ 3.65.
Therefore, in  Gaia-Enceladus  the fraction of \livii\ rich giants is of about 1\%, very close to the 1-2\% found in the Galaxy. \citet{Casey2019}   argue that Li rich giants  are likely binary systems in the red clump. However, \citet{adamow2018}  monitored a sample of 15 Li-rich giants within  a Planet-Search program and found a normal binary  fraction.
The \citet{Casey2019}   suggestion still requires a full observational investigation, but  it would be  interesting to investigate  if the Li-rich giant found in Gaia-Enceladus is also a binary system.

    \section{Beryllium abundances in Gaia-Enceladus}
    
  \beix\   is  burnt in the interior of  the stars and  is made   through  Galactic Cosmic Ray  (GCR)  spallation reactions  in the  interstellar medium  \citep{Reeves1970,Meneguzzi1971A&A}.  Energetic cosmic rays  with energies $>$ 100 MeV  hit CNO atoms at rest in the ambient interstellar gas and break them into smaller pieces, producing   Li, Be and B.     A reverse  mechanism is working also during  supernovae  explosions which accelerate  nuclei of  C, N, and O  which later collide with protons and $\alpha$ particles in the surrounding medium  and  break up  into  
smaller units.  The only suitable   transitions of beryllium are the \beixii\ resonance lines which  fall  at 313.0 nm close to the atmospheric cut off. For this reason  \beix\  is  a very challenging element to be measured  in Galactic halo stars   \citep{Molaro1984,Molaro1997,Smiljanic2009,Boesgaard2011}.  In extragalactic stars it  will probably remain  out of reach also for the next generation of giant 40 m class telescopes.  However,
  few stars belonging to Gaia-Enceladus have measured  \beix\  and arguably these could  be arguably considered as the first extragalactic  \beix\ measurements.

\begin{figure}
	\includegraphics[width=\columnwidth]{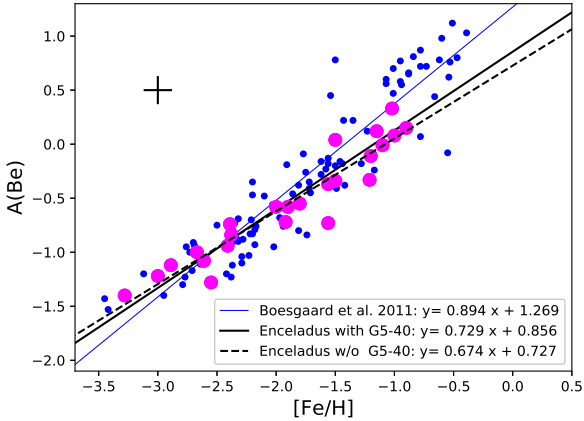}
    \caption{
    A(Be) abundances versus iron abundances. 
   A(Be) abundances  from \citet{Boesgaard2011} are in blue dots. The    Gaia-Enceladus star candidates are highlighted in magenta, with a squared symbol for G5-40. The cross on the top left corner shows  the mean errors in the abundances reported in Table \ref{tab:literature}. The solid and dashed black lines are   the best fit through the Gaia-Enceladus stars with and without G5-40, respectively. The solid blue line is the best fit through the   \citet{Boesgaard2011} data points without Gaia-Enceladus candidates. }
    \label{fig:beryllium}
    \end{figure}

\begin{figure}
	\includegraphics[width=\columnwidth]{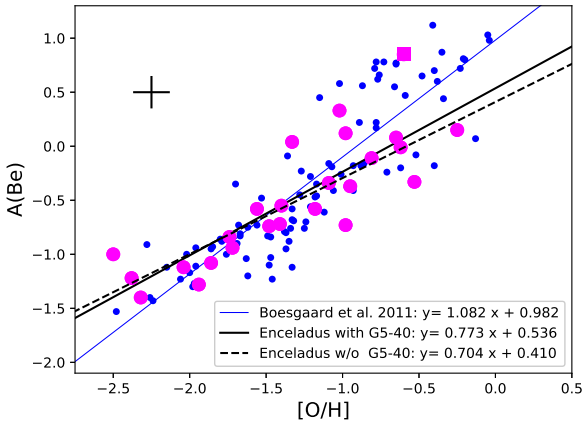}
    \caption{
   A(Be) abundances versus oxygen abundances. Symbols and lines are the same as in Figure \ref{fig:beryllium}. }
    \label{fig:beo}
    \end{figure}

\begin{figure}
	\includegraphics[width=\columnwidth]{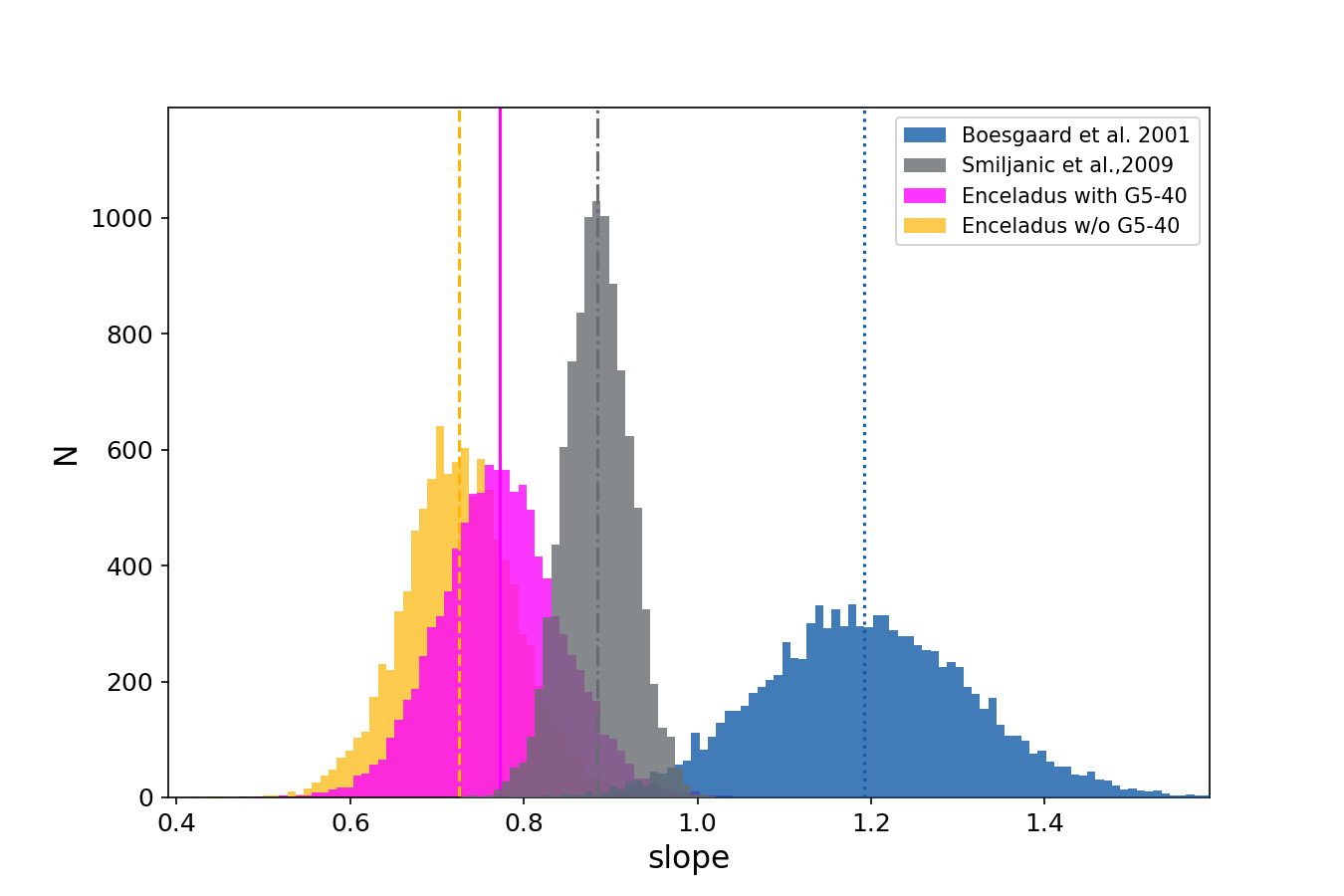}
    \caption{ Distribution of the slopes  from  a linear fitting with a  MCMC of 10000 chains.
    The median slope value of each data set is marked with vertical line:
   blue dotted line for  \citet{Boesgaard2011}, grey dash-dot line  for \citet{Smiljanic2009}, magenta solid line
    for Enceladus with G5-40,
    and yellow dashed line for Enceladus without G5-40.
    }
    \label{fig:beta}
    \end{figure}

 These  25  stars  are listed in Table \ref{tab:literature} and are shown in Figure \ref{fig:beryllium} together with the Galactic measurements.   The \beix\ abundance in these  stars shares the same location of the Galactic stars but lay preferentially at   lower \beix\ abundances for a given metallicity. 
Figure \ref{fig:beryllium} shows the relationship between [Fe/H] and A(Be) for the Gaia-Enceladus stars together with the data points from \citet{Boesgaard2011}.
We  found that the linear fit between
these two logarithmic quantities for  the Gaia-Enceladus stars is:

\begin{equation}
A(Be) =   0.729 (\pm 0.059)  [Fe/H] + 0.856 ( \pm 0.117)
\end{equation}

 When G5-40 is not considered  slope and intercept  become 0.674$\pm 0.048 $ and 0.727$ \pm 0.098$, respectively with a dispersion of 0.16 dex. The regression found  for the Gaia-Enceladus candidate stars  is significantly  different than that   found with 
\citet{Boesgaard2011}  data points. After taking out   the 17 data points in common with  Gaia-Enceladus sample, the remaining  98 measurements provide  the relation:
\begin{equation}
A(Be) = 0.894(\pm 0.041)[Fe/H] + 1.269 (\pm 0.078)
\end{equation}

The two slopes and intercepts differ by 2.3 $\sigma$,  3.5 without G5-40,  and   2.9 (3.5) $\sigma$, respectively. However, it seems that this difference is mainly produced by the Galactic A(Be) measurements for   [Fe/H]> -1.  
\citet{Smiljanic2009} found an even  steeper  slope but with  a smaller number  of very metal-poor stars.  For      stars with [Fe/H $>-$2.2, the  slope is of 1.04 $\pm$0.06 in \citet{Boesgaard2011} 
and  1.16 $\pm$0.07 found by \citet{Smiljanic2009}.

 To asses with confidence that we  deal with two different populations, we have performed a MCMC
\citep[following the method described in][]{mcmc}
of the A(Be)  vs [Fe/H] by taking into account the errors reported in Table \ref{tab:literature}. The MCMC result with 10000 chains for the Gaia-Enceladus values  is 
$A(Be) =  0.770(\pm0.070) [Fe/H] + (0.957\pm0.140) $ 
The slope is slightly flatter if star G5-40 is not considered:
$A(Be) =  0.725(\pm0.065) [Fe/H] + (0.845\pm0.133) $.
While the MCMC fitting for the \citet{Boesgaard2011} and \citet{Smiljanic2009} samples, once cleaned from the Enceladus candidate stars,   are 
$A(Be) = 0.884(\pm0.038)[Fe/H] + 1.245(\pm0.075) $ 
and $A(Be)= 1.186(\pm0.121) [Fe/H] + 1.587(\pm0.126) $, respectively.  
In  Figure \ref{fig:beta} are shown the  distribution of the slope values of the Gaia-Enceladus candidates, the \citet{Boesgaard2011}  and \citet{Smiljanic2009} stars.
The slope of the Gaia-Enceladus candidates stars, no matter if  star G5-40 has been taken into account, is flatter than that of \citet{Boesgaard2011} and \citet{Smiljanic2009}.
We apply a two-side K-S statistic to check the significance of the slope difference.
The result shows a zero  probability that the MCMC slope distribution between the Gaia-Enceladus stars and the \citet{Boesgaard2011, Smiljanic2009}  are similar.
However,  the main cause of the difference  with  \citet{Smiljanic2009} is  probably due to the fact that they have much fewer stars at low metallicity.

\begin{figure}
	\includegraphics[width=7.5truecm]{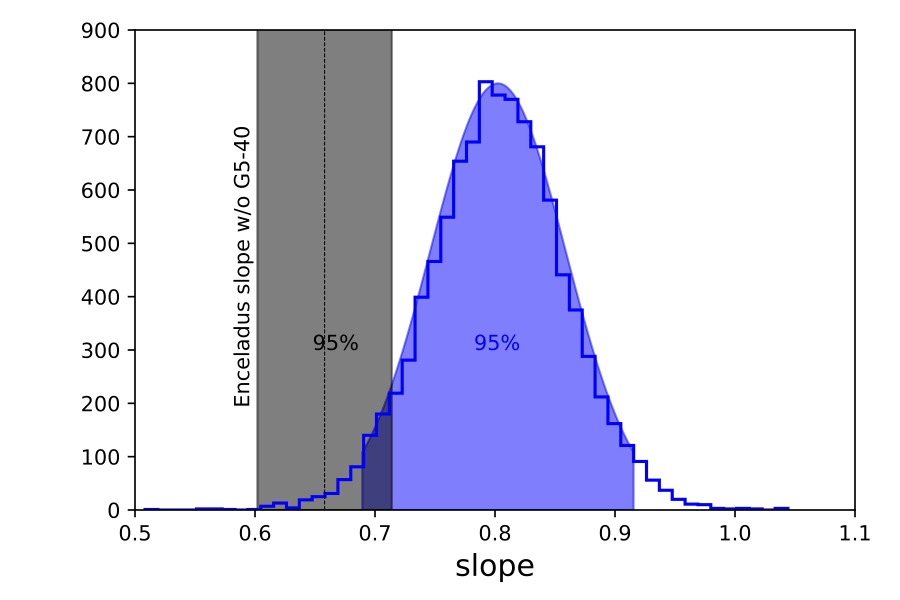}
    \caption{  Distribution of slopes for 10000 samples of 25 stars randomly drawn from the 115 stars of \citet{Boesgaard2011}. The slope refers to the family of lines  passing through  the common  origin: [Fe/H]=$-$2.85, A(Be)=$-$1.17. The slope of the  Gaia-Enceladus star candidates without G5-40 is also shown.}
    \label{fig:distribution}
    \end{figure}

We note that there is  considerable overlap of the A(Be) abundances at the low metallicity end between Gaia-Enceladus and the MW. For instance, the average of 6 stars in the Gaia-Enceladus sample with metallicity [Fe/H]$<$-2.2  provides  $<A(Be)>$ = -1.18 at $<[Fe/H]>$ =-2.83. In the \citet{Boesgaard2011}  there are 15 stars in this metallicity range with $<A(Be)>$= -1.17 and $<[Fe/H]>$ = -2.86.  Since they share  the same mean values at low metallicities, to evaluate a different growth we considered  the lines passing through it A(Be)= -1.17 + a([Fe/H]+2.85), and fitted the remaining points of the two data samples. The results are     a=0.695($\pm 0.034$), or  a=0.658($\pm 0.028$) without G5-40, for Gaia-Enceladus and  a=0.825($\pm 0.023$) for the \citet{Boesgaard2011} cleaned sample. The two slopes differ by  3.17, or 4.6  without G5-40 $\sigma$,  and confirm the different  growth  in the two populations.
To further test that this result is not casual  we took 10000  samples of 25 stars randomly drawn from the sample of 115 stars in \cite{Boesgaard2011} sample. The  distribution of the  slopes is provided  in Figure \ref{fig:distribution}. They show   a    mean of a=0.802 $\pm 0.056$  differing by 2.3 $\sigma$  from that of Gaia-Enceladus  without G5-40.

 As mentioned before, \beix\  production is directly related to oxygen rather than to  iron. Therefore, we show in Fig.\ref{fig:beo} the   available determinations for beryllium and oxygen reported in Table \ref{tab:literature}. We have also calculated  the correlations for Gaia-Enceladus:
\begin{equation}
A(Be) = 0.536 (\pm 0.105)  [O/H] + 0.773(\pm 0.153) 
\end{equation}

with a  dispersion of  0.30, or  A(Be) = 0.385($\pm 0.139$) + 0.700($\pm 0.0937$)  [O/H] without G5-40, with a dispersion of 0.26 dex.
The  analysis of the remaining 98 data points in  \citet{Boesgaard2011}  provides:

\begin{equation}
A(Be) = 1.082(\pm 0.0642)[O/H] + 0.982 (\pm 0.0854) 
\end{equation}

 with a dispersion of 0.36. Although the data points are slightly more scattered, probably due to  the difficulty of the oxygen determination, the regression analysis  shows  a  flatter  and less scattered slope for the Gaia-Enceladus candidates stars in comparison with  that of the Milky Way.

\citet{Smiljanic2009} found    statistical evidence for an intrinsic scatter in the halo stars in the A(Be)-[Fe/H] relation, above what is expected from observational errors. The observed scatter in the Galaxy is of the order of 0.5 dex. 
while  the dispersion of the \beix\ abundances in Gaia-Enceladus stars is of  0.19 dex, which is comparable to the measurement errors. For  comparison the dispersion of the data point along the fit of the 98 data points in \citet{Boesgaard2011} is  0.298 dex, i.e. almost a factor two larger than  the data of the Gaia-Enceladus candidates.
The  uncertainties in the  A(Be) abundance determination comes from  the atmospheric 
parameters, mainly the log g,  and from  the location  of the pseudo-continuum and unidentified blends.  The total uncertainty   is of $\approx$ 0.15 dex, which is of the same order of the observed dispersion in Gaia-Enceladus stars.

Overall the A(Be) versus [Fe/H] behaviour of Gaia-Enceladus seems to belong to a very homogeneous stellar population with a very smooth beryllium enrichment. On the other hand the scatter observed in the Galaxy could  be originated  by the presence of multiple stellar populations   with  different time scales in the \beix\  evolution  as has been also suggested by \citet{Smiljanic2009}.

 \citet{Rich2009} found that the dependence of A(Be) on [Fe/H]  shows distinct differences in  the accretive group of Galactic stars. The latter   show a flatter   slope of A(\beix) with  [Fe/H]  than the Galactic stars. 
They ascribe this different behaviour  to the differing importance of the two mechanisms for \beix\  formation, i.e.  in the vicinity of SN II stars or 
preferentially by GCR spallation reaction. 
\citet{Rich2009}  found that  the accretive and retrograde groups show  
A(Be) $\le$ 0.35 and the star   G21-22 with A(Be)=0.31 is the highest value.  We note that G 5-40, the most metal rich star in our sample of Gaia-Enceladus candidates, shows A(Be)=0.85. However,  this star is a bit out to the general trend in the Be-metallicity plot. Thus, it would be of interest if a similar cut found for the accretive stars applied also to Gaia-Enceladus.  This would be possible   by searching for \beixii\ in  other metal rich stars of the Gaia-Enceladus  galaxy.  As we have seen  omitting G 5-40 from the Gaia-Enceladus bona fide candidates makes the difference between Gaia-Enceladus and the Galaxy  more significant.

\begin{figure}
\includegraphics[width=\columnwidth]{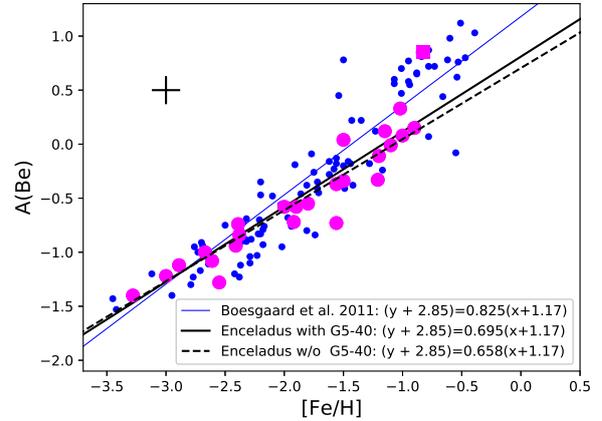}
    \caption{  Best fit for Gaia-Enceladus and  Milky Ways data points for the family of lines  passing through  the common  origin: [Fe/H]=$-$2.85, A(Be)=$-$1.17. Symbols as in Figure \ref{fig:beryllium}.
    }
    \label{fig:comp}
    \end{figure}

\section{Discussion}

Among the light Big Bang elements  D and \ivhe\ are observed in  extragalactic objects. Most of the \ivhe\ measures come from the local universe at
redshifts typically of 0.01 to 0.1 at most while  Deuterium is observed up to a
redshift of 3 or even 4.  \iiihe\  can be measured only in the Galaxy \citep{Bania2010} and the  same applies to \livii\, since main sequence solar type  stars in external
galaxies are out of reach even for the 10m class telescopes. Upper limits
for the interstellar Li towards the SN1987A in the LMC were obtained when the supernova was as bright as V$\approx$ 4 mag \citep{Molaro1989}. A  detection of interstellar \livii\ has been reported by \citet{Howk2012} in the SMC in the line of sight towards  the star SK143.    

\citet{Molaro1997}  suggested that Li could have  been detected in Galactic
stars, which might possibly have been born in other galaxies.
\citet{Preston1994} in their HK objective prism survey of metal poor
stars in the Milky Way identified a population of stars which they called the blue metal poor main sequence stars (BMP). This population is composed by hot and metal poor objects that should have already evolved from the main sequence if coeval with the
halo stars. The space density for the BMP stars is about one order of magnitude larger than that of blue stragglers in globular cluster, thus suggesting that
field BS are a minor component of this population. Moreover, the kinematical properties of the BMP are intermediate among those of halo and thick-disk populations. In \citet{Preston1994}, the authors suggested  that the BMP population has been accreted from a low luminosity satellite of the Milky Way in the recent past. One of these stars is CS 22873-139 which has a remarkably low metallicity [Fe/H]=-3.1. This object is a spectroscopic binary for which   \citet{Preston1994}  derived  an  age of  $<$ 8 Gyr, which again is too short for an  halo star. Li abundance  in star CS 22873-139  has  the canonical halo value of
A(Li) = 2.28. 

Recently several efforts have been made to measure {\it extragalactic} \livii\ abundance. Omega Centauri is  a globular cluster-like stellar system characterized by a wide range of metallicities and probably of ages, and probably was 
stripped from the core of a dwarf galaxy. \citet{Monaco2010}
found that  $\Omega $ Cen dwarfs display a constant Li abundance
observed among the stars  spanning a wide range of ages and metallicities that overlap  with the Spite plateau.   \citet{Mucciarelli2014} by means of  stellar modeling have been able to derive  the initial lithium abundance in the  globular cluster  M54 in the nucleus of the Sagittarius dwarf galaxy.
The Sagittarius galaxy  is at 25 Kpc  and the main sequence stars  are of 22 mag and   too faint to be studied at high resolution. The only possibility are  Red Giant Branch stars where  the \livii\ abundance  has been modified by a post MS dilution.    By considering  dilution \citet{Mucciarelli2014}  have established an initial Li abundance of this stellar system (A(Li)= 2.29$\pm$ 0.11  or  2.35$\pm$ 0.11 dex, when accounting also for atomic diffusion.

 The analysis of the Gaia-Enceladus stars confirms the discrepancy between the primordial nucleosynthesis prediction and the metal poor stars of this dwarf galaxy suggesting that   the Li cosmological problem is ubiquitous and present also in other galaxies, regardless of their type. Thus,
a solution able to explain the discrepancy must work both
in the Milky Way and  for other galaxies, with likely experienced different origins
and star formation histories. As noted by \citet{Mucciarelli2014},  it is  unlikely that the scenario proposed by \citet{Piau2006}, requiring    one third of the gas in the Galactic halo    re-processed by Population III  massive stars, could  be valid also   in a smaller galactic system such as    Gaia-Enceladus. 

For a    chemical evolution model  which assume that  lithium is mainly produced by novae  it is expected that a dwarf galaxy like  Gaia-Enceladus the Li abundance  will rise from  the Spite plateau at a  metallicity lower than  in  the Galactic thin disk.  
Future observations targeting  hot dwarf stars of Gaia-Enceladus with metallicity in the range $\approx$ -1.5, -1.0  will allow to establish the presence of a  slightly different Li evolutionary behaviour.

The formation of \beix\   occurs 
during supernovae explosions when CNO atoms accelerate out into the
ambient gas and  strike protons and neutrons  
splitting  into smaller atoms. 
An interesting feature is the fact that the \beix\   abundances of Gaia-Enceladus stars  starting from the same origin  at the lowest metallicities, then show a more gentle rise and the tendency to populate the lower region of the Galactic halo stars. A possible explanation is that SNeIa   contributed to the  abundance of iron of the Gaia-Enceladus  stars. The relative contribution of SNe Ia to  Fe is higher in a low star formation galaxy as Gaia-Enceladus compared to the Milky Way. This would produce an enhancement in iron without a corresponding \beix\ production and, therefore,     a smaller \beix\ abundance in Gaia-Enceladus for a given metallicity is expected.  However, if this is the case the Be-O relations  should not differ. Alternatively,  there could be   a steeper increase of A(Be) in the Galaxy for  [Fe/H] $\ge$ -1.5 possibly due to an increase in  importance  of the inverse spallation process at high metallicity. This phase is marginally   seen in Enceladus  and the two relations could  appear different when fitting onto the whole metallicity range.

\beix\ is only produced by spallation of cosmic rays and its abundance allows  to
constrain the degree to which $^7$Li, $^6$Li
$^{11}$B and $^{10}$B  may have been
produced by the same processes. Thus,
 the amount of \livii\, and \livi\,  produced by spallation of high energy cosmic rays in the Gaia-Enceladus can be inferred from the observed Be by taking the 
ratio of the cross sections of spallation processes  ($^7$Li)/$^9$Be=7.6 and ($^6$Li)/$^9$Be =5.5 \citep{Steigman1992}.  Observationally it is not possible to resolve  $^6$Li   from $^7$Li and they both are  considered  contributing to the Li 670.7 nm line. 
However, $^6$Li is rather fragile 
and is not expected to survive in halo stars where it is not detected. Therefore, we consider only    the $^7$Li/$^9$Be relative cross section, providing 7.6 \citep{Molaro1997}.
The predicted fraction of Li produced by spallation processes in Gaia-Enceladus should follow a relation with similar slope but shifted by 0.88 dex.  The low  \beix\ abundance in Gaia-Enceladus implies a relatively small  contribution by spallation processes to the Li observed in Gaia-Enceladus stars.   
These abundances are virtually extragalactic measurements of \beix\  and therefore hold a more general significance. 

Besides standard galactic cosmic rays, it has been suggested that additional production of light elements might come from cosmic rays accelerated in galaxy-galaxy interactions \citep{Prodanovic2013}.  In support of this possibility are produced the high lithium values in the ISM  of M82 \citep{Ritchey2015} and the high \livi/\livii/ ratio in the SMC of \citep{Howk2012}. Within a simplified framework \citet{Prodanovic2013} showed  that large-scale tidal shocks from a few galactic fly-bys can possibly produce light elements in amounts comparable to those expected from the interactions of galactic cosmic-rays produced in supernovae over the entire history of a system. These effects are particularly evident for dwarf galaxies. In the case of the SMC, they  found  that only two such fly-bys could  account for as much lithium as the standard  galactic cosmic ray production channel.
  The same processes should lead to an additional amount of \beix\ and in 
  particular in dwarf galaxies  such as Gaia-Enceladus which suffered the tidal collision that resulted into the merging with the Galaxy.  The observations of Gaia-Enceladus do not reveal any excess of \beix\  compared to the Galaxy and restrict  the possibility of an additional \beix\ nucleosynthetic channel.
  However, in case of  a  head-on collision  with a  direct merger  no extra production of the light elements is expected.

\section{Acknowledgements}
   We  thank Alexandro Saro for his advice regarding statistical tests and acknowledge useful discussions with   Tijana Prodanovic on  the effects of tidal production of light elements. We warmly thank an anonymous referee for several suggestions which improved  the paper significantly. 
  The use of TOPCAT tool \citep{topcat},  the Simbad database and the VizieR catalog access tool of CDS \citep{cds}  are also acknowledged. This work has been partially supported by the the EU COST Action CA16117 (ChETEC).  XF acknowledges the support of the National Natural Science Foundation of China under grant number 11973001 and National Key R\&D Program of China No. 2019YFA0405504.  The GALAH survey is based on observations made at the Australian Astronomical Observatory, under programmes A/2013B/13, A/2014A/25, A/2015A/19, A/2017A/18. We acknowledge the traditional owners of the land on which the AAT stands, the Gamilaraay people, and pay our respects to elders past and present.

\bibliographystyle{mnras}
\bibliography{biblio_litio}

 \begin{table*}
\centering
\caption{ Kinematical properties of the selected Gaia-Enceladus candidates. The energy (En) and the  angular momentum in the z direction  (Lz)  are from  \citet{Helmi2018}.  
The apocenter distance (R$_{APO}$), the pericenter distance (R$_{PER}$),  the max distance from the Galactic plane (Z$_{max}$) and the  eccentricity (Ecc), are computed  from  the stellar orbit of the past 1 Gyr.}
\begin{tabular}{lccccccc}
\hline
\multicolumn{1}{c}{Gaia source\_id}  &
\multicolumn{1}{c}{Name}  &
\multicolumn{1}{c}{Energy}  &
\multicolumn{1}{c}{Lz}  &
\multicolumn{1}{c}{r$_{apo}$}  &
\multicolumn{1}{c}{r$_{peri}$}  &
\multicolumn{1}{c}{ecc}  &
\multicolumn{1}{c}{Zmax}  \\
\hline
    6086864760409366656  &  TYC 8248-1737-1  &    -156076.36  &       -183.31  &         16.07  &          0.72  &          0.91  &          0.93  \\ 
    4725550450463451904  &  L  126-11        &    -143328.41  &         73.65  &         20.79  &          0.24  &          0.98  &          1.13  \\ 
    5459976109889190144  &  TYC 7174-224-1   &    -177744.13  &       -332.32  &         10.57  &          1.09  &          0.81  &          0.27  \\ 
    5750434405835685888  &  --               &    -166586.73  &        -32.53  &         12.74  &          0.12  &          0.98  &          7.17  \\ 
    6679323239394561792  &  --               &    -160505.95  &       -552.18  &         14.46  &          1.37  &          0.83  &          1.60  \\ 
    5946574193490564480  &  --               &    -137327.40  &        -33.12  &         22.11  &          0.24  &          0.98  &         18.50  \\ 
    5781595596159463040  &  TYC 9429-2667-1  &    -179647.78  &      -1113.11  &          8.52  &          3.69  &          0.40  &          1.73  \\ 
    6383892436469819008  &  --               &    -129193.33  &         47.69  &         27.30  &          0.09  &          0.99  &          9.18  \\ 
    6729270234418615552  &  CD-38 13129      &    -129202.75  &       -108.51  &         27.57  &          0.38  &          0.97  &          6.07  \\ 
    5242632244811706496  &  TYC 9213-2091-1  &    -165717.38  &       -325.14  &         13.08  &          1.14  &          0.84  &          1.98  \\ 
    3155410389590889856  &  G  89-14         &    -168977.85  &       -176.82  &         12.65  &          0.50  &          0.92  &          0.85  \\ 
      32655224762711936  &  G4-36            &    -141767.94  &        114.54  &         20.51  &          0.09  &          0.99  &         14.55  \\ 
    3846427888295815552  &  HE0938+0114      &    -123889.65  &       -260.90  &         27.75  &          0.73  &          0.95  &         24.80  \\ 
     866863321051682176  &  BD+24 1676       &    -133448.07  &        -77.43  &         25.70  &          0.34  &          0.97  &          0.47  \\ 
    1289512635833404032  &  G166-47          &    -178769.02  &       -133.59  &          9.83  &          0.54  &          0.90  &          5.01  \\ 
    4761346872572913408  &  HIP 24316        &    -157275.95  &       -869.57  &         14.78  &          2.47  &          0.71  &          6.55  \\ 
    5181063205724188032  &  G75-56           &    -170575.00  &      -1466.88  &          9.94  &          5.06  &          0.33  &          1.65  \\ 
    1776289248313154688  &  BD+17 4708       &    -149027.24  &       -201.64  &         18.62  &          0.55  &          0.94  &          0.77  \\ 
    2658240166703766016  &  BD+02 4651       &    -172793.13  &       -626.12  &         11.27  &          1.65  &          0.75  &          1.61  \\ 
    4272653983123701120  &  G21-22           &    -163379.73  &         36.43  &         13.94  &          0.17  &          0.98  &          1.18  \\ 
    2910503176753011840  &  LTT2415.00       &    -121743.08  &      -2497.55  &         31.97  &          6.71  &          0.65  &          9.74  \\ 
     125750427611380480  &  G37-37           &     -95894.36  &       -410.31  &         51.67  &          1.25  &          0.95  &         28.61  \\ 
      29331710349509376  &  G05-19           &    -140895.94  &       -219.78  &         21.66  &          0.64  &          0.94  &          7.00  \\ 
    2905773322545989760  &  HIP 25659        &    -114045.58  &      -3228.30  &         39.49  &          7.99  &          0.66  &          2.48  \\ 
    5586241315104190848  &  HD59392          &    -178043.45  &       -640.99  &         10.32  &          1.68  &          0.72  &          0.47  \\ 
    5551565291043498496  &  CD-48 2445       &    -164908.62  &       -275.84  &         13.62  &          0.82  &          0.89  &          0.14  \\ 
     949652698331943552  &  G107-50          &    -167967.85  &       -777.61  &         12.60  &          2.00  &          0.73  &          0.45  \\ 
    5617037433203876224  &  W 0725-2351      &    -161949.71  &       -587.79  &         11.12  &          2.31  &          0.66  &         10.26  \\ 
     870628736060892800  &  G88-10           &    -178734.94  &       -650.69  &          8.68  &          2.38  &          0.57  &          5.33  \\ 
    1097488908634778496  &  G234-28          &    -161952.45  &       -428.06  &         13.14  &          1.22  &          0.83  &          6.85  \\ 
    6268770373590148224  &  HD 140283        &    -159638.76  &       -139.94  &         14.80  &          0.51  &          0.93  &          1.00  \\ 
     731253779217024640  &  HIP 52771        &    -173197.37  &       -547.17  &         10.80  &          1.68  &          0.73  &          4.68  \\ 
     791249665893533568  &  BD+51  1696      &    -167411.87  &        -64.93  &         12.75  &          0.33  &          0.95  &          1.09  \\ 
    2279933915356255232  &  BD+75   839      &    -172679.06  &        -37.69  &         11.70  &          0.31  &          0.95  &          1.09  \\ 
    2427069874188580480  &  HIP 3026         &    -177036.14  &        -80.12  &         10.72  &          0.39  &          0.93  &          1.37  \\ 
    5486881507314450816  &  HIP 34285        &    -146177.34  &         53.60  &         19.59  &          0.15  &          0.98  &          1.56  \\ 
    6859076107589173120  &  HIP 100568       &    -178122.94  &        -63.71  &         10.33  &          0.35  &          0.93  &          4.62  \\ 
    4468185319917050240  & {\bf  G 20-24}          &    -169181.12  &       -209.53  &         12.24  &          0.70  &          0.89  &          4.55  \\ 
    4376174445988280576  &  BD +2 3375       &    -129172.93  &        141.57  &         27.43  &          0.25  &          0.98  &          0.52  \\ 
     588856788129452160  &  BD 9 2190        &    -127301.32  &      -1813.78  &         28.25  &          4.97  &          0.70  &          8.46  \\ 
     61382470003648896  & G 5-40 & -174809.98 & -28.78 & 10.64   &   0.35    &    0.94     &    6.37 \\
    4715919175280799616  &  HIP 7459         &    -149799.68  &       -303.19  &         18.36  &          0.72  &          0.92  &          0.12  \\ 
    3643857920443831168  &  G64-37           &    -152525.63  &      -1058.08  &         15.01  &          3.08  &          0.66  &          8.64  \\ 
    5709390701922940416  &  HIP 42592        &    -168527.88  &       -608.18  &         12.16  &          1.61  &          0.77  &          3.31  \\ 
    5184824046591678848  &  LP 651-4         &    -172739.89  &         98.15  &         11.16  &          0.02  &          1.00  &          6.15  \\ 
     761871677268717952  &  BD 36 2165       &    -156655.26  &       -289.70  &         15.13  &          0.93  &          0.88  &         10.57  \\ 
    2722849325377392384  &  BD 7 4841        &    -176665.28  &        -44.94  &         10.87  &          0.24  &          0.96  &          0.23  \\ 
      16730924842529024  &  BD 11 468        &    -177010.67  &       -939.28  &         10.01  &          2.73  &          0.57  &          0.71  \\ 
    1458016709798909952  &  BD 34 2476       &    -142582.26  &         27.03  &         17.50  &          0.21  &          0.98  &         18.73  \\ 
    5806792348219626624  &  CD -71 1234      &    -122305.09  &         23.77  &         29.48  &          0.15  &          0.99  &         26.74  \\ 
    3699174968912810624  &  HE1208-0040      &    -174470.01  &       -652.29  &         10.57  &          1.86  &          0.70  &          1.15  \\ 
    4228176122142169600  &  G 24-25          &    -167077.60  &       -511.44  &         12.50  &          1.35  &          0.81  &          3.32  \\ 
    5133305707717726464  &  BD -17 484       &    -147101.71  &       -284.96  &         18.96  &          0.84  &          0.91  &          5.18  \\ 
\hline
\end{tabular}
\label{tab:cinematica}
\end{table*}

\begin{table*}
\centering
\caption{Online Table: Enceladus Candidates from the GALAH survey with Li determinations. In the second column the flag Cannon from GALAH is reported. A flag Cannon = 0 means a reliable measurement \citep{Buder18}}.
\label{tab:galah}
\begin{tabular}{lccccc}
\hline
\multicolumn{1}{c}{Gaia source\_id}  &
\multicolumn{1}{c}{f. c.}  &
\multicolumn{1}{c}{\teff}  &
\multicolumn{1}{c}{\logg}  &
\multicolumn{1}{c}{[Fe/H]}  &
\multicolumn{1}{c}{A(Li)}  \\
\hline
    1732135056069825920  &      0  &    5033.32  &       2.64  &      -1.62  &       1.10  \\ 
    1753089552970858112  &      0  &    4509.55  &       0.85  &      -1.17  &      -0.06  \\ 
    2682935850698314240  &      0  &    5048.96  &       2.71  &      -1.77  &       1.29  \\ 
    2708796085010318592  &      0  &    5016.32  &       2.47  &      -0.86  &       1.34  \\ 
    2712879572412229504  &      0  &    4836.28  &       2.59  &      -0.48  &       1.11  \\ 
    2891158957586832384  &      0  &    4778.17  &       1.92  &      -1.25  &       1.20  \\ 
    2894294794812883072  &      0  &    5331.82  &       2.53  &      -1.02  &       1.42  \\ 
    2894411381703462400  &      0  &    5077.36  &       2.31  &      -1.08  &       1.32  \\ 
    2955975885301234176  &      0  &    5055.64  &       2.62  &      -0.93  &       1.43  \\ 
    2967759557578771712  &      0  &    4878.00  &       2.06  &      -0.63  &       0.57  \\ 
    2988442509461752704  &      0  &    5275.19  &       2.44  &      -0.84  &       1.27  \\ 
    3155073114396829696  &      0  &    5142.80  &       2.84  &      -1.79  &       1.34  \\ 
    3155410389590889856  &      0  &    5834.23  &       3.80  &      -1.32  &       2.32  \\ 
    3198867766340862464  &      0  &    5201.73  &       2.88  &      -1.30  &       1.44  \\ 
    3287575368035203456  &      0  &    4924.48  &       2.15  &      -1.51  &       1.06  \\ 
    3549336066900457856  &      0  &    4725.84  &       1.73  &      -0.91  &       0.67  \\ 
    4345499789558948608  &      0  &    4863.51  &       1.93  &      -1.63  &       1.13  \\ 
    4382559791045030912  &      0  &    5130.73  &       3.05  &      -1.75  &       1.27  \\ 
    4422595036636105984  &      0  &    5129.61  &       2.49  &      -1.05  &       1.40  \\ 
    4488499007696406784  &      0  &    4929.01  &       2.32  &      -0.87  &       1.13  \\ 
    4637675213429717376  &      0  &    4687.05  &       2.56  &      -0.50  &       0.72  \\ 
    4640304072716065280  &      0  &    4966.53  &       2.33  &      -0.92  &       1.39  \\ 
    4641720351067505024  &      0  &    5279.22  &       2.43  &      -1.13  &       1.57  \\ 
    4646028100186033152  &      0  &    4420.65  &       1.94  &      -0.34  &      -0.18  \\ 
    4685472457068177536  &      0  &    5119.64  &       2.77  &      -1.71  &       1.29  \\ 
    4693690589220392192  &      0  &    5581.52  &       3.64  &      -0.81  &       1.84  \\ 
    4699007101601586304  &      0  &    4739.82  &       1.78  &      -0.98  &       1.17  \\ 
    4701711045508666112  &      0  &    4904.03  &       2.08  &      -1.48  &       1.22  \\ 
    4705815178817583360  &      0  &    5087.22  &       2.25  &      -0.73  &       1.37  \\ 
    4725550450463451904  &      0  &    5857.36  &       3.79  &      -1.02  &       1.87  \\ 
    4775145228104639232  &      0  &    5221.97  &       3.06  &      -1.66  &       1.35  \\ 
    4881266547068247552  &      0  &    5060.26  &       2.79  &      -1.59  &       1.33  \\ 
    5225454471572336384  &      0  &    4656.60  &       1.46  &      -1.44  &       0.70  \\ 
    5226653836902677888  &      0  &    5119.06  &       2.30  &      -1.02  &       1.41  \\ 
    5242632244811706496  &      0  &    5985.94  &       3.90  &      -1.30  &       2.31  \\ 
    5374060481756321024  &      0  &    5056.90  &       2.68  &      -0.58  &       1.20  \\ 
    5384620324564238720  &      0  &    5125.24  &       2.36  &      -0.92  &       1.29  \\ 
    5389369703698845056  &      0  &    4625.94  &       1.85  &      -0.98  &       1.03  \\ 
    5395064211857452672  &      0  &    5097.59  &       3.16  &      -0.76  &       1.40  \\ 
    5401796104940236928  &      0  &    4988.28  &       2.35  &      -1.06  &       1.33  \\ 
    5458455798843435264  &      0  &    4855.07  &       2.12  &      -0.86  &       1.24  \\ 
    5459976109889190144  &      0  &    5929.12  &       4.03  &      -1.12  &       1.85  \\ 
    5470192432632884480  &      0  &    4982.57  &       2.72  &      -1.95  &       1.11  \\ 
    5668871503711755008  &      0  &    5156.47  &       2.87  &      -1.24  &       1.31  \\ 
    5673500722542380032  &      0  &    4905.31  &       2.03  &      -0.63  &       0.67  \\ 
    5676472530674105856  &      0  &    5090.21  &       2.47  &      -0.90  &       1.46  \\ 
    5728832958015766016  &      0  &    5063.25  &       2.58  &      -1.01  &       1.39  \\ 
    5729276503582050304  &      0  &    4707.00  &       1.70  &      -0.82  &       1.01  \\ 
    5750434405835685888  &      0  &    5834.30  &       3.84  &      -1.16  &       2.35  \\ 
    5752994756100253440  &      0  &    4969.32  &       2.39  &      -1.55  &       1.31  \\ 
    5781595596159463040  &      0  &    5935.60  &       3.92  &      -1.21  &       1.96  \\ 
    5791415819848472960  &      0  &    5122.25  &       2.72  &      -1.60  &       1.29  \\ 
    5807140176154682368  &      0  &    4773.03  &       1.79  &      -1.79  &       1.02  \\ 
    5816364872565566592  &      0  &    4823.02  &       1.97  &      -1.96  &       0.92  \\ 
    5818897872469627264  &      0  &    4961.09  &       1.79  &      -0.95  &       1.06  \\ 
    5946574193490564480  &      0  &    5981.59  &       3.83  &      -1.17  &       2.28  \\ 
    6002366471495563392  &      0  &    4969.72  &       2.34  &      -0.54  &       0.87  \\ 
    6002820050086770944  &      0  &    4458.54  &       0.98  &      -1.91  &       0.39  \\ 
    6014472781696679296  &      0  &    4837.48  &       1.97  &      -0.37  &       2.56  \\ 
    
\hline
\end{tabular}
\end{table*}

\begin{table*}
\centering
\caption{Continue: Enceladus Candidates from the GALAH survey with Li determinations.}
\label{tab:galah1}
\begin{tabular}{lccccc}
\hline
\multicolumn{1}{c}{Gaia source\_id}  &
\multicolumn{1}{c}{f. c.}  &
\multicolumn{1}{c}{\teff}  &
\multicolumn{1}{c}{\logg}  &
\multicolumn{1}{c}{[Fe/H]}  &
\multicolumn{1}{c}{A(Li)}  \\
\hline
    6069724164405817856  &      0  &    4870.25  &       2.36  &      -0.65  &       0.66  \\ 
    6086864760409366656  &      0  &    5901.98  &       4.06  &      -0.90  &       2.32  \\ 
    6094877043381522688  &      0  &    5129.83  &       2.84  &      -1.19  &       1.48  \\ 
    6145484020253118208  &      0  &    5016.55  &       2.49  &      -1.37  &       1.30  \\ 
    6195538909150136960  &      0  &    5159.91  &       2.85  &      -1.30  &       1.52  \\ 
    6195538909150136960  &      0  &    5159.91  &       2.85  &      -1.30  &       1.52  \\ 
    6322264015162573952  &      0  &    5093.28  &       3.24  &      -0.55  &       1.17  \\ 
    6353051474611711232  &      0  &    5087.26  &       2.69  &      -1.65  &       1.27  \\ 
    6353984380164650624  &      0  &    4710.40  &       1.52  &      -1.27  &       1.00  \\ 
    6379689896869892480  &      0  &    4879.05  &       2.12  &      -1.12  &       1.26  \\ 
    6380579470496738688  &      0  &    5164.42  &       3.03  &      -1.68  &       1.41  \\ 
    6383892436469819008  &      0  &    5740.34  &       4.05  &      -1.23  &       2.19  \\ 
    6391140554558850432  &      0  &    5018.87  &       2.35  &      -1.27  &       1.46  \\ 
    6392042218516940416  &      0  &    5230.59  &       3.10  &      -1.80  &       1.33  \\ 
    6399437190128061824  &      0  &    4921.87  &       2.15  &      -1.35  &       1.35  \\ 
    6402012766053572096  &      0  &    5282.14  &       3.25  &      -1.53  &       1.48  \\ 
    6403691685950215808  &      0  &    4778.33  &       1.74  &      -1.56  &       1.20  \\ 
    6404161074334487808  &      0  &    4906.98  &       2.27  &      -1.35  &       1.26  \\ 
    6410480414335655680  &      0  &    4911.41  &       2.70  &      -0.67  &       0.84  \\ 
    6460648101957895040  &      0  &    4888.83  &       2.10  &      -1.35  &       1.37  \\ 
    6470581639759021056  &      0  &    4921.61  &       2.28  &      -1.26  &       1.38  \\ 
    6478310489244756224  &      0  &    4925.78  &       2.21  &      -1.15  &       1.39  \\ 
    6481444784579188352  &      0  &    5149.27  &       2.98  &      -1.84  &       1.36  \\ 
    6483276601013988864  &      0  &    5212.00  &       2.31  &      -0.88  &       1.12  \\ 
    6487462686594876544  &      0  &    4703.52  &       1.63  &      -0.92  &       0.60  \\ 
    6495347039663212800  &      0  &    4841.89  &       1.93  &      -1.31  &       1.15  \\ 
    6562694673781993472  &      0  &    5029.37  &       2.23  &      -0.98  &       1.23  \\ 
    6565548112614542592  &      0  &    5008.35  &       2.15  &      -1.24  &       1.33  \\ 
    6649405764224188416  &      0  &    5077.31  &       2.23  &      -0.62  &       1.13  \\ 
    6679323239394561792  &      0  &    5888.48  &       4.07  &      -1.16  &       2.23  \\ 
    6685311175422378368  &      0  &    5257.84  &       3.09  &      -1.61  &       1.43  \\ 
    6690424007572533120  &      0  &    4841.04  &       2.02  &      -1.17  &       0.65  \\ 
    6717997250935351680  &      0  &    4793.81  &       1.88  &      -1.38  &       1.15  \\ 
    6729270234418615552  &      0  &    5955.72  &       3.96  &      -1.24  &       2.01  \\ 
    6789958195324077824  &      0  &    4977.80  &       2.25  &      -1.25  &       1.12  \\ 
    6837496641407476224  &      0  &    5242.97  &       2.95  &      -1.13  &       1.52  \\ 
    6891710810994077568  &      0  &    4779.46  &       2.78  &      -0.39  &       0.64  \\ 
    3154778621377895936  &      1  &    5405.12  &       2.72  &      -1.06  &       1.64  \\ 
    3174650919659325952  &      1  &    5659.52  &       3.80  &      -1.43  &       1.97  \\ 
    3838992131675444480  &      1  &    5222.12  &       3.88  &      -1.26  &       1.29  \\ 
    4297160826005222656  &      1  &    6035.02  &       3.43  &      -1.04  &       2.00  \\ 
    4379282249962462336  &      1  &    5553.20  &       4.28  &      -1.21  &       1.54  \\ 
    4705330358613974144  &      1  &    5282.20  &       4.24  &      -1.23  &       1.25  \\ 
    4710644401389442048  &      1  &    5716.63  &       3.08  &      -1.00  &       1.65  \\ 
    4820909925710430976  &      1  &    5342.10  &       3.20  &      -1.79  &       1.43  \\ 
    5203339380025535104  &      1  &    5462.45  &       3.04  &      -1.06  &       1.79  \\ 
    5671614613424216064  &      1  &    5437.16  &       2.82  &      -1.09  &       1.64  \\ 
    5806792348219626624  &      1  &    6071.23  &       3.84  &      -1.27  &       2.26  \\ 
    5817597047142573952  &      1  &    5182.77  &       3.91  &      -1.49  &       1.44  \\ 
    6180667945665056768  &      1  &    5442.22  &       3.39  &      -1.61  &       1.68  \\ 
    6359046634778502144  &      1  &    5713.37  &       3.14  &      -1.20  &       2.02  \\ 
    6562024315286365568  &      1  &    5884.12  &       3.71  &      -1.20  &       2.12  \\ 
    6580611245819304320  &      1  &    5567.00  &       3.03  &      -1.18  &       1.56  \\ 
    6630376452061688704  &      1  &    5336.72  &       3.03  &      -0.98  &       1.74  \\ 
    6697722432614775040  &      1  &    5606.67  &       2.91  &      -0.92  &       1.66  \\ 
    5419728418033942528  &      2  &    4218.52  &       0.73  &      -0.78  &      -0.61  \\ 
    5947512381830659328  &      2  &    5043.14  &       2.89  &      -1.57  &       1.32  \\ 
    6682809550247700096  &      2  &    6026.64  &       3.94  &      -1.36  &       2.10  \\ 
    5607538065058737536  &      3  &    4711.13  &       0.89  &      -0.45  &       0.49  \\ 
    6694628127719929088  &      3  &    5017.92  &       1.69  &      -0.96  &       0.57  \\ 
    6198863767893020288  &      8  &    5198.20  &       3.02  &      -1.54  &       1.22  \\ 
    6095297464840155008  &     64  &    4843.43  &       2.15  &      -1.25  &       1.22  \\ 
    6381173172415485568  &     65  &    4834.19  &       3.92  &      -1.75  &       0.92  \\ 
\hline
\end{tabular}
\end{table*}

\bsp	
\label{lastpage}
\end{document}